\documentclass[conference]{IEEEtran}
\IEEEoverridecommandlockouts
\makeatletter
\def\ps@headings{%
\def\@oddhead{\mbox{}\scriptsize\rightmark \hfil \thepage}%
\def\@evenhead{\scriptsize\thepage \hfil \leftmark\mbox{}}%
\def\@oddfoot{}%
\def\@evenfoot{}}
\makeatother
\usepackage{cite}
\usepackage{amsmath,amssymb,amsfonts}
\usepackage{amsthm}
\usepackage{enumerate}
\usepackage{booktabs}
\usepackage{algorithmic}
\usepackage{graphicx}
\usepackage{textcomp}
\usepackage{xcolor}
\usepackage{verbatim}
\usepackage{color, colortbl}
\def\BibTeX{{\rm B\kern-.05em{\sc i\kern-.025em b}\kern-.08em
    T\kern-.1667em\lower.7ex\hbox{E}\kern-.125emX}}

\definecolor{lgreen}{rgb}{0.84, 1, 0.88}

\begin{document}



\title{Long Passphrases: Potentials and Limits}


\author{\IEEEauthorblockN{Christopher Bonk}
\IEEEauthorblockA{\textit{Ontario Tech University}\\
Oshawa, Canada \\
chris@chrisbonk.ca}
\and
\IEEEauthorblockN{Zach Parish}
\IEEEauthorblockA{\textit{Ontario Tech University}\\
Oshawa, Canada \\
zachary.parish@ontariotechu.net}
\and
\IEEEauthorblockN{Julie Thorpe}
\IEEEauthorblockA{\textit{Ontario Tech University}\\
Oshawa, Canada \\
julie.thorpe@ontariotechu.ca}
\and
\IEEEauthorblockN{Amirali Salehi-Abari}
\IEEEauthorblockA{\textit{Ontario Tech University}\\
Oshawa, Canada \\
abari@ontariotechu.ca}
}

\maketitle

\begin{abstract}
Passphrases offer an alternative to traditional passwords which aim to be stronger and more memorable. However, users tend to choose short passphrases with predictable patterns that may reduce the security they offer. To explore the potential of long passphrases, we formulate a set of passphrase policies and guidelines aimed at supporting their creation and use. 
Through a 39-day user study we analyze the usability and security of passphrases generated using our policies and guidelines. Our analysis indicates these policies lead to reasonable usability and promising security for some use cases, and that there are some common pitfalls in free-form passphrase creation. Our results suggest that our policies can support the use of long passphrases.  
\end{abstract}


\section{Introduction}
Passwords are the dominant form of knowledge-based authentication to protect resources such as email accounts, banking credentials, and even critical infrastructure. Despite their ubiquity, the security flaws inherent to passwords are well known. Users often select passwords which are easy to guess, especially when attacks are targeted. Policies designed to improve password security, such as length, digit and symbol requirements can be effective in increasing the difficulty of cracking a password, but often negatively impact usability\cite{Herley2009, Inglesant2010, Komanduri2011}. When password systems have particularly poor usability, users can be prompted to engage in behaviors such as password reuse and recording, which further harm security. 

While many attempts have been made to replace text passwords\cite{Bonneau-Cormac}, they have endured as the most common means of authentication. This may be due to the simplicity of text-based authentication, the familiarity that most users have with them, and their ease of deployability\cite{Bonneau-Cormac}. Password managers allow users to store stronger passwords without the need to remember them \cite{Bonneau-Cormac}; however, they still require the user to recall a strong master password and it is often advised that password managers do not store high-risk accounts (e.g., email and financial)\cite{ITSAP2019}. Thus, the need for users to remember at least a few strong passwords remains.

Passphrases seek to provide an alternative text-based authentication method that can leverage the simplicity and user familiarity of passwords, while providing a higher degree of security and memorability\cite{porter1982}. With passphrases, users create a sequence of words to use as their secret, rather than a shorter string of characters. In principle, the longer length of passphrases provides additional security by increasing the search space of an attack. In practice, users have been observed to create short passphrases which consist of easily guessable sequences of words, often closely following natural language \cite{Bonneau2012}. While this similarity to natural language may improve the usability of passphrases systems\cite{Danescu-Niculescu-Mizil2012}, 
it reduces the theoretically large search space and simplifies an attacker's task. This motivates our research to explore long passphrases that aim for stronger security, while retaining reasonable memorability. 

To explore the limits and potential of long passphrases, we craft a set of passphrase policies that encourages users to select strong passphrases.  To promote memorability, we draw upon research on human memory to inform our policies and guidelines to support the creation of long passphrases.  
We study the usability of our resulting passphrase policies in a 39-day user study.  Our results suggest that our policies and guidelines supported users in generating long and possibly more secure passphrases. 

Our contributions include:
\begin{enumerate}[(i)]
\item The design of a set of policies and guidelines for supporting users' creation of long passphrases.
\item A usability analysis of passphrases created under our policies and guidelines, based on results from a multi-session 39-day user study.
\item A security analysis of the passphrases created in our user study.
\item A discussion of major pitfalls in passphrase creation that were revealed from our analysis.
\end{enumerate}

\section{Related Work}
We focus on research in passphrases, both in the classic sense where a phrase structure is required, and also passwords with longer length requirements.

%
\vspace{10pt}
\noindent\textit{Security of Long Passwords.} Passwords with longer length requirements have been shown to provide comparable security when contrasted with shorter passwords with more complex character requirements. In their user-study of nearly three-thousand participants, Komanduri et al. \cite{Komanduri2011} found that only ~1\% of passwords generated with a 16-character length requirement could be guessed, as compared to 0\% for passwords with the requirement of 8-character and a minimum of one uppercase, lowercase, digit and symbol characters (also known as comp8). Both outperformed passwords with only an 8-character length requirement, of which ~19\% were guessed \cite{Komanduri2011}.

\vspace{10pt}
\noindent\textit{Passphrase Security.}  The security provided by passphrases however is less well studied, often only with rough estimates provided by potential character combination calculations \cite{Keith2007}, which have been suggested to provide an inaccurate view of security \cite{Bonneau2012} as users will not chose character combinations at random. 
Empirical analysis has shown that when users create passphrases of shorter lengths, they are vulnerable to dictionary-based attacks due to predictable patterns in user choice \cite{Bonneau2012}. This mirrors results from the study of mnemonic passwords, which are based on a phrase, where underlying mnemonics have been shown to come from easy to predict sources\cite{Kuo2006}. Shay et al. \cite{Shay2014} found that a policy requiring a minimum length of 16 characters and at least two words (i.e., letter sequences separated by non-letter sequences), when compared to the comp8 policy, provided comparable security in limited online attacks (0.3\% vs 0.1\% respectively) and considerable improvement against longer offline attacks (6.6\% vs 23.5\% respectively). 
The passphrases generated by their users under the 2-words and 16-characters policy had a median length of 18 characters and contained three words rather than the required two 31.8\% of the time.

\vspace{10pt}

\noindent\textit{Passphrase Usability.}
In comparisons with traditional passwords, passphrases have demonstrated comparable or improved usability metrics \cite{Komanduri2011,Keith2007,Shay2014}, though users can experience a learning curve\cite{Keith2007}. The memorability of passphrases (e.g., two-word requirement with a 16 character requirement) has been shown to be comparable to that of passwords generated under the comp8 policy, with slightly lower rates of storage (51.3\% vs 56.9\%), and a similar number of average login attempts (1.4 vs 1.4) \cite{Shay2014}. Keith et al. \cite{Keith2009} found that users experienced fewer memory errors using passphrases (3 to 5 word sequences with at least 16 characters) when compared to passwords (8-characters with at least one non-letter). 
They also found that users with passphrases that are similar to written language had a lower typographic error rate than users with randomly generated passwords (2.34\% vs 5.36\%) \cite{Keith2009}. Despite these memorability improvements, the length of passphrases has been shown to increase typographical errors and therefore decrease login success rates \cite{Keith2007} when compared to passwords (71.58\% successful logins vs 85.61\%)
, an effect which worsens with passphrase length \cite{Shay2012}. Error correction schemes have been proposed as a solution to this problem \cite{Bard2007, Keith2007}, allowing users to achieve similar login rates with passphrases as with passwords  (85.86 vs 87.50\%). 
In their user study on geographic hints for passphrases, Addas et al. \cite{addas2019} found an absolute recall rate of 25\%, markedly lower than the previously discussed studies; however, this study involved the creation and recall of 4 different passphrases. 
Woo and Mirkovic \cite{woo2016} applied mnemonics to guide the creation of passphrases and serve as passphrase hints. Users created their passphrases based on a randomly generated mnemonic and were cued with it at login time. Their results showed an increase in recall from 40\% to 69.6\% after 7 days and a reduction of common phrase use from 51\% to under 5\% when mnemonics were applied to user generated passphrases.

\vspace{10pt}
\noindent\textit{System-Assigned Passphrases.}
While our work focuses on user-generated passphrases, previous work on the memorability of system-assigned passwords and passphrases are of interest. System-assigned passphrases can offer a guaranteed amount of security, since the passphrases are drawn randomly from a pool of possible ones. However, most studied systems have drawbacks in one way or another. Bonneau and Schechter \cite{Bonneau} performed a two-week remote user study of system-assigned 56-bit codes. Users began by entering an 18.8 bit code (e.g., 2 words or 4 letters), which was extended over the study to the full 56-bit size. 88\% of users could recall  their code from memory after 3 days and 94\% by the end of their study. These positive results suggest users have the ability to remember longer authentication strings; however, only 59\% of users could recall their code after a two week break, suggesting that frequency of use is key. Shay et al. \cite{Shay2012} compared the usability of system assigned passphrases to passwords of similar entropy. Discouragingly, they found 72\% of participants stored their assigned secret and that only 48.5\% of users who did not store their secret could recall it between 2 and 5 days after assignment. System-assigned secrets using visual, verbal or spacial cues \cite{Ameen2015} have shown considerable improvement in recall when compared to other work. Recent work explored allowing users some choice in composing their passphrase by having them select words from a set of random words drawn from a dictionary. This guided word-choice method achieved 97-99\% of the maximal entropy of randomly generated passphrases, while increasing recall from 23.08\% to 40.43\% and 50.98\% for sets of size 20 and 100 respectively \cite{blanchard2018}. Other work has applied implicit learning techniques to aid in learning system-assigned passphrases \cite{joudaki2018}, which was shown to significantly improve recall rates and login times when compared to a control group after 7-8 days; however, the passphrases studied with this technique were low entropy (20 bits).

\section{Designing Policies and Guidelines} \label{sec:policy-design}
We draw upon existing research in human memory and previous authentication systems to inform our policy design. 
We first present the rationale for our design choices and then our final policies (requirements) and guidelines (recommendations).

\vskip 2mm
\noindent\textit{Passphrase Length.}
The human mind can recall 5--7 \emph{chunks} of information in short term memory \cite{Baddeley1994, Cowan2001, Miller1956}. To strive for the longest passphrases in this range, we suggest a length requirement of at least 7 words. 


\vskip 2mm
\noindent\textit{Use of Proper Noun.}
To promote the use of word choices that have more lexical distinctiveness and are thus more memorable \cite{Danescu-Niculescu-Mizil2012}, we require users to use at least one proper noun in their passphrase.  Given the relatively large number of proper nouns (e.g., at least 39,336 unique city names\cite{MaxMind} and 33,542 unique person names\cite{SSA2013}), this may also increase an attacker's search space.


\vskip 2mm
\noindent\textit{Avoid Common Subphrases.}
To mitigate attacks that exploit language trends (e.g., \cite{Bonneau2012}), we suggest that users avoid using the top ten-thousand n-grams (for n=3, 4, or 5) in their passphrase.\footnote{Results from preliminary testing of a passphrase system blacklisting the top 10,000 n-grams revealed a severe decrease in usability when n=2. Therefore, we suggest only blacklisting common sequences of 3, 4 or 5 words.} We suggest this be positioned as a recommendation, since pilot testing found users had difficulty interpreting it as a requirement. This suggestion is based on our n-gram analysis of the Corpus of Contemporary American English (COCA)\cite{Davies}, comprised of 450 million words of American English text derived from spoken, fiction, non-fiction, magazines, newspapers, and academic research. This analysis found that the top 10,000 covered the most highly frequent n-grams, thus should reduce the chance of highly common phrases being chosen. 
\vskip 2mm
\noindent\textit{Passphrase as a Story.}
Memory recall is improved when ideas are related to one another\cite{Craik1972}, or memories are approached as a story\cite{Carminatti2012}. For this reason, we suggest recommending users formulate their passphrase as a story.

\vskip 2mm
\noindent\textit{Use Slang and Non-Dictionary Words.}
To promote the use of uncommon word choices, we recommend users to use slang words in their passphrases. The goal of this recommendation is to increase the likelihood of rare words, that may be difficult to enumerate from an attacker's perspective. Also, as slang can be used in different ways, an attacker might have further difficulty to enumerate candidate passphrases that employ them. For example, UrbanDictionary.com\cite{AaronPeckham2014} contains 681,981 unique slang terms. 

\vskip 2mm
\noindent\textit{Use a Mnemonic.} Mnemonics can offer memorability improvements for passwords \cite{Juang2012}. For this reason, we suggest building a step to guide users to create a mnemonic directly into the system to encourage their use. \\



Our design choices can be summarized as a passphrase policy and set of user guidelines:  
\vskip 2mm
\noindent \textbf{Passphrase Policy.} \emph{The passphrase is required to
\begin{itemize}
\item{Contain at least 7 words, separated by spaces}.
\item{Contain at least one proper noun (i.e., names of people, places, etc.)}.
\end{itemize}
}


\vskip 2mm
\noindent \textbf{User Guidelines.} \emph{To improve security and memorability, we recommend users to
\begin{itemize}
\item {Use slang or non-dictionary words (e.g., `bazinga').}
\item {Not employ common three word phrases (e.g.,`all of it').}
\item{Formulate the passphrase as a story.}
\item{Select a mnemonic.}
\end{itemize}
}

\section{Data Collection}
We conducted a hybrid online/in-person user study, split into 4 sessions and held across 39 days, to assess our policies and guidelines.

\vskip 1.5mm
\noindent \textit{System Implementation.} 
Our implementation, that we named \emph{StoryPass}, presents users with our user guidelines and policies. To encourage using a story and mnemonic, we build separate steps into the system.  More specifically, prior to passphrase creation, users were instructed to make a short one-to-two sentence story to build their passphrase from. Users then create their passphrase given the policies and first two recommendations listed above. After a user creates a passphrase, they are instructed to type a mnemonic word to aid in remembering their passphrase. 
Users are told that their mnemonic can be anything that will aid in memory (e.g., objects like business cards, or a particular memory). This step aims to improve memory by guiding the user to memorize their passphrase using an anchor or association. To mitigate the typographic errors associated with long passphrases \cite{Keith2009}, we accept an entered passphrase as valid when the Levenshtein distance between the input and original passwords is at most $0.125$. This equates to a single error per eight characters and is consistent with previous work \cite{Keith2009}. We also normalize capitalization to lowercase and discard punctuation.

\vskip 1.5mm
\noindent \textit{Session 1 (Day 1, in-lab).} Users were instructed to create a short story of one-to-two sentences and then a word mnemonic that serves as a reminder of their passphrase (e.g., an object, idea, or a memory).  They practiced creating passphrases and mnemonics until they were able to successfully confirm. Users then created the study passphrase and mnemonic they used for the rest of the sessions. 
Users next completed a demographic questionnaire which also included information about their computer skills and password habits. Finally, users logged in using their study passphrase.


\vskip 1.5mm
\noindent \textit{Session 2 (Day 2--3, online).}
Session 2 was held one day after Session 1 to simulate user self-reported frequency of logging into email accounts \cite{Hayashi2011}. Users were asked to login with their passphrase. 
After three unsuccessful login attempts, users were given the option to reset their passphrases. After 10 failed attempts, users were required to reset. For these users, an extra Session 2 was held the following day and for the rest of the study, these users completed each session one day later than other participants.

\vskip 1.5mm
\noindent \textit{Session 3 (Day 9--10, in-lab)}
Session 3 was held in-lab 7 days after Session 2 to simulate user reported frequencies of logging into financial websites (e.g., banks) \cite{Hayashi2011}. Users were asked to login by their passphrase with the same process as Session 2. 
We also conduct the post-study questionnaire in this session as we anticipated significant dropout for Session 4.

\vskip 1.5mm
\noindent \textit{Session 4 (Day 39--40, online)}
We held Session 4 one month after Session 3 to simulate the user reported average frequency of logging into some e-commerce websites. Users were asked to login by their passphrase. This session also follows the same format as Session 2. 

\vskip 1.5mm
\noindent \textit{Demographics.} We recruited 40 participants through a mass email and on-campus posters in our university.  All 40 participants completed Sessions 1 and 2. However, Session 3 and 4 were completed by 35 and 31 of those participants, respectively.  One participant who completed the study experienced a technical glitch that led us to exclude their results.
There were 21 males and 19 females. Participant ages ranged from 18 to 62 (mean=34.2). Participant education levels ranged from high-school to masters degrees with the mode of high-school diploma. 
The first language of participants were English (47.5\%), Urdu (12.5\%), Chinese (10\%), Spanish (7.5\%), Persian Farsi (5\%), and others (17.5\%).

\section{Results}
\begin{figure}
\begin{center}
\includegraphics[scale=.45]{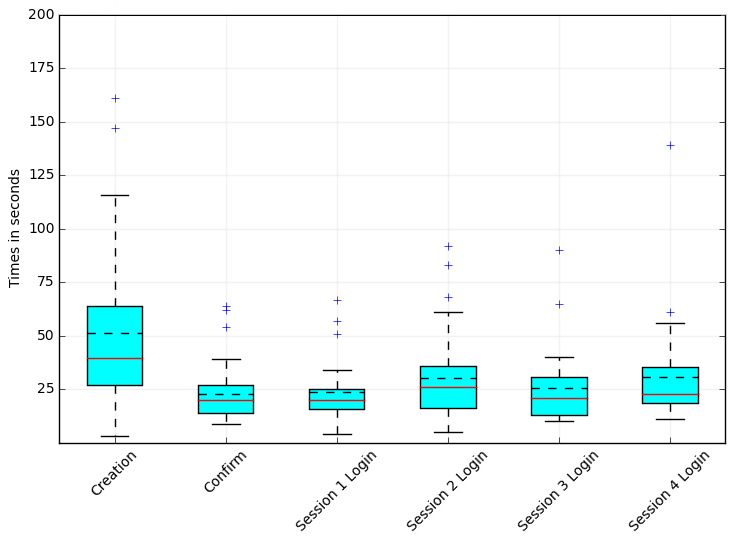}
\vspace{-12pt}
\caption{The passphrase creation, confirm, login times for all sessions.}
\label{fig:TimeAnalysisFig}
\end{center}
\end{figure}

We perform 
a usability analysis of the passphrases created using our policies and guidelines.
We analyze the creation time, login time, login success rate, password recording, failed login attempts, and passphrase resets for user in our study. We also analyze the impact of error correction and the average edit distance of each successful login.
\vskip 1.5mm
\noindent \textit{Creation and Login Time.} We measure passphrase creation time and all session login times to identify if users had difficulty with these tasks. We exclude failed login attempts so that times are not affected by technical error (e.g., a user pressing enter before they were done typing). 
Figure \ref{fig:TimeAnalysisFig} shows the time users took to create, confirm, and successfully login with their passphrase during each session. The creation time has an average of less than 40 seconds. The confirmation and login times for all four sessions have consistently an average below 30 seconds.  

\vskip 1.5mm
\noindent \textit{Login Success Rate.}
A measure of memorability, the login success rate is the percentage of successful logins compared to all login attempts. Table \ref{tab:loginrates} shows the login success rate for each session. Users had more difficulty logging in during Session 1 and 2 than Session 3. This could be the result of users becoming more familiar with passphrases by Session 3. Success rates were slightly lower in Session 4 than that of Session 3, possibly due to the long period between these two sessions.


\begin{table}
\centering
\caption{The login success rates for each session.}
\label{tab:loginrates}
\begin{tabular}{ccccc}
\toprule
Session 1 & Session 2 & Session 3 & Session 4\\
\midrule
77\% & 74\% & 86\% & 77\%\\
\bottomrule

\end{tabular}
\end{table}


\vskip 1.5mm
\noindent \textit{Failed Logins.}
We also recorded each user's cumulative failed login attempts up to Sessions 3 and 4. Out of 35 participants who completed Session 3, 24 had no failed login attempts, 5 had 1--3, 4 had 4--10 and 2 had more than 10. Out of the 31 participants who completed Session 4, 20 had no failed login attempts, 5 had 1--3, 3 had 4--10 and 3 had more than 10.  While the majority of users had no difficulty logging in, a small group exhibited great difficulties, accumulating over 10 failed logins each. Motivated by this, we performed further analysis of the failed logins.

%

Our analysis revealed two major passphrase creation pitfalls which contributed to failed logins: (1) using two many common words which could be easily changed or reordered. For example, ``today my friends went shopping at the mall'' could be mistaken as ``my friends went shopping at the mall today''; (2) using a series of random words, (i.e., having no grammatical structure or logical relationship).
Another significant cause of failed logins 
was attempting to enter the passphrase without spaces. It is possible that previous habits using passwords caused these login failures, suggesting additional instruction may be necessary.

\vskip 1.5mm
\noindent \textit{Error Correction and Edit Distances.}
Table \ref{tab:passphraseErrorCorrection} shows that the average Levenshtein distance over successful login attempts for all sessions are relatively low ($< 2\%$). Through further analyses, we discovered that some failed logins were caused by typographical errors. On successful login attempts, users most often correctly recalled their passphrase within an edit distance of one.  55 successful logins from Session 1 to 4 had edit distance of greater than zero, with the maximum of 11.9\%. Only 3 logins had an edit distance over 10\%. 65\% of successful login attempts (36 of 55) had an edit distance less than 6.125\%, half of the maximum error tolerance. 18 of 39 users who returned for Session 3 said or were observed being surprised at the system accepting their passphrase as a correct login. This can be attributed to passphrases being significantly longer and more complex than normal or regularly-used passwords and users not being aware of the error tolerance.


\begin{table}
\centering
\caption{Average Levenshtein distance (\%) over successful logins.}
\label{tab:passphraseErrorCorrection}
\begin{tabular}{cccc}
\toprule
Session 1 & Session 2 & Session 3 & Session 4\\
\midrule
0.9 & 1.0 &1.8 & 1.7\\
\bottomrule

\end{tabular}
\end{table}

\vskip 1.5mm
\noindent \textit{Passphrase Resets.}
We also recorded the number of passphrase resets for all sessions. Of 39 total users who began the study, 33 never reset their passphrases, 4 participants had one reset and 2 participants had two resets.

\vskip 1.5mm
\noindent \textit{Storage.} During in-lab sessions (Sessions 1 and 3), we observed if participants referred to any form of storage to help remember their passphrase. Our Session 3 questionnaire also asked participants if they wrote down their passphrase. Of 39 users, only 15 users (38\%) stored their passphrases. 

\vskip 1.5mm
\noindent \textit{Mnemonics.}
During passphrase creation, we instructed users to create a mnemonic to aid in passphrase memorability. Users were not reminded of their mnemonic until Session 3, when we asked users to recall their mnemonic and report if they wrote it down. 26 of 35 participants were successful in remembering their mnemonic, with 28 of 35 participants reporting they did not write it down. 
Although many participants had forgotten that they had made a mnemonic during Session 1, the majority were successful in remembering it during Session 3.

\vskip 1.5mm
\noindent \textit{Usability Questions.} 

During Sessions 1 and 3, users completed a pre and post-study questionnaire, respectively. 
In general, users seemed to like using passphrases: 52\% agreed (20\% disagreed) that they ``could easily use this method of logging in every day.''   Some users reported the length requirement of a passphrase to be too large than a traditional password. ``I would use StoryPasses for all of my accounts'' and ``I would not use StoryPass for any of my accounts'', are both strongly disagreed with, suggesting users see a potential role for this method of authentication, but not for all of their accounts. Most users perceived that their StoryPass was more secure than a traditional text password. 
Users indicated that they would not use StoryPass for infrequently used accounts and agreed that they could remember their passphrases for up to one year. Many users indicated verbally and in the additional comments section that they enjoyed using passphrases and planned to apply the StoryPass guidelines for future passphrases.

\vskip 1.5mm
\noindent \textit{Potential Usability Impact of Blacklisting 2-grams.}
Our pilot testing lead us to avoid blacklisting popular 2-grams for usability reasons. To assess and reconfirm such a policy decision further, we searched the generated passphrases for the top ten-thousand 2-grams. 31 of the 39 (80\%) passphrases had at least one 2-gram from the top 10,000. This implies that if we would have blocked 2-grams, 80\% of users would have faced some  difficulties in selecting their passphrases.  Allowing these 2-grams is therefore likely a necessity for usability.

\vskip 1.5mm
\noindent \textit{Users with Unusual Difficulties.}
Four users had great difficulties with passphrases across usability metrics. These users had average login rates of 27\% compared to 88\% for all other users and had on average 12.3 failed logins as compared to 0.7 for all other users. Memory errors were the most common error for these users, either entering a completely incorrect passphrase, partial passphrase, or entering the correct words in an incorrect order. Some of these users also attempted to enter their passphrase without spaces. One common memory issue was to mix up the order of nouns, pronouns and conjunctions. For example, ``Everyone has been bowling at Lucky Strike this week'' could be entered as ``Everyone this week has been bowling at Lucky Strike.'' Users also experienced difficulties when their passphrase did not contain proper grammatical structure (e.g., ``Pear top montreal green up down middle''). 

One participant had a passphrase that suffered from neither of our pitfalls, but still experienced difficulties. This user 
rated their computer skills as the lowest possible score. While they required much more time to practice the creation and had many failed login attempts in Session 1, they had a perfect login record for the rest of the study. This suggests that users with a low level of computer skills may require additional time and instruction to use passphrases effectively.

\vspace{10pt}

\section{Discussion}
We discuss a comparison of our system with other related work, a number of findings of particular interest, and limitations herein.

\subsection{Usability Comparison}

\begin{table*}
\centering
\caption{Comparisons between the usability measures of our work and previous work}
\label{tab:comparison}
\begin{tabular}{lccccc}
\toprule
Password/Passphrases  & Login Time & Success Rate & Login Attempts & Storage Rate & Reset Rate \\ 
\midrule
comp8 \cite{Komanduri2011}  & n/a & n/a & n/a & 50.0\% & 11.1\% \\ 
basic16 \cite{Komanduri2011} & n/a & n/a & n/a & 33.0\% & 11.1\% \\ 
password \cite{Keith2009}  & n/a  & 75.91\% & 1.3 & n/a & n/a \\ 
passphrase \cite{Keith2009} & n/a  & 90.93\% & 1.1 & n/a & n/a \\ 
comp8\cite{Shay2014} & 13.2 s & n/a & 1.4  & 56.9\% & 15.5\% \\ 
basic20 \cite{Shay2014} & 15.3 s & n/a & 1.3 & 50.0\% & 15.5\% \\ 
2word16\cite{Shay2014} & 14.5 s & n/a & 1.4  & 51.3\% & 15.5\% \\ 
UPass\cite{woo2016}  & n/a & 40.0\%--52.3\%  & n/a & n/a & n/a \\  
UPassHint\cite{woo2016}  & n/a & 69.6\%--71.4\% & n/a & n/a & n/a\\ 
GeoHints\cite{addas2019} & 100 s & 25\% & 3.3 & n/a & n/a \\ 
\textbf{StoryPass (ours)} & 25--30 s  & 74\%--86\% & n/a  & 38.5\% & 15.4\%\\ 
\bottomrule
\end{tabular}
\end{table*}

Table \ref{tab:comparison} provides a comparison between the usability of our policies and guidance vs.\ related work on passphrases and passwords. While these studies have differences in methodology, it provides a basis for what we should expect from the usability of our StoryPass implementation. 
\emph{Login Time} shows the average login time in seconds. \emph{Success Rate} shows the percentage of successful login attempts (over all login attempts). \emph{Login Attempts} shows the number of attempts before a successful login.  \emph{Storage Rate} is the percentage of users who stored/recorded their login information in some way. \emph{Reset Rate} is the percentage of users who required a reset or reminder of their login information. 
The main take-away from this comparison is that the metrics for StoryPass are within the same range as other password and passphrase systems, with the exception of login time, which seems to be about twice as long as password systems.  The reason for this is clear: a 7-word passphrase is much longer (and thus hopefully more secure than) an 8-20 character password.  This usability/security tradeoff is important when considering which context a passphrase system is appropriate.

\subsection{Major Pitfalls in Passphrase Creation}
Our analysis highlights the need for semantic and lexical structure to aid memory and typographical load. Participants who used passphrases that had a grammatically correct structure and consistent underlying message, like a story, fared much better at successfully logging in. Improved usability appears to come from the use of memory aiding factors such as sentence-like structure or choosing a distinct and personalized idea relevant to the user. 
We identified two major pitfalls in passphrase creation that lead to decreased usability: (1) using common words which could be easily changed or reordered, and (2) using a series of random words without grammatical structure. For our system to be applied successfully users should be made aware of these pitfalls and instructed against them. Similar patterns were noted by Addas et al. \cite{addas2019}, who reported that 56.1\% of failed logins resulted from inexact recall of the exact word order or choice of words. 

\subsection{Security Analysis}
To evaluate guessability of the passphrases collected in our study, we tested them against the frequency-ranked list of the most common 492,630 words from the Corpus of Contemporary American English (COCA)\cite{Davies}, which is derived from 450 million words of American English text derived from spoken, fiction, non-fiction, magazines, newspapers, and academic research.  If a given passphrase has all component words in this list, we mark it as `guessable' using this corpus.  Our results indicate that  59\% of the passphrases would be considered guessable by this metric. How long it might take an attacker to guess these passphrases (containing 7 or more words) depends on how they choose to order guesses. The most naive method would be to brute-force all possible combinations of 7-word passphrases from the COCA dictionary. This method is rendered impractical by the attack dictionary size of $492630^7\approx 2^{132.37}$. A more plausible approach is to chain n-grams from COCA, such that the last word of a preceding n-gram overlaps with the first word of the following n-gram (e.g., ``international conference" could be chained with ``conference on"). The generated n-gram chains can be ordered based on the frequency of each component n-gram. To test how well this strategy might work, we tested how many passphrases whose words would have been fully matched to any set of (overlapping) n-grams in the COCA corpus. Our results showed that none (0\%) of the passphrases from our study would be guessed with this strategy. This shows that even when our participants chose somewhat common words, the way they ordered them was not very common. However, we caution against claiming these passphrases are secure against language modeling attacks, as other strategies may be possible based on other natural language processing techniques.

Analysis of the collected passphrases also revealed that the context of the passphrase creation is relevant when considering security. $\frac{3}{39}$(7\%) of passphrases created had a strong correlation to our institution and included terms such as professor names, the institution name, or the name of the study administrator. Additionally, $\frac{2}{39}$(5\%) of the passphrases were found on Google searches of the exact passphrase, so they may be vulnerable to passphrase dictionaries compiled from common long phrases.

\subsection{Limitations}
The high number of students in our study may result in higher average computer skills in our sample, improving security and usability results. We attempted to mitigate this by excluding users from a computer security background.

We took a number of steps to ensure that our collected results are as authentic as possible. We conduct sessions 1 and 3 of the study in a controlled room designed for administering user studies, during which only two principle investigators were involved to provide consistent instructions and record observations. The same script was used by investigators in both in-lab sessions. This lab environment however, may have altered user's behaviors, an issue we attempted to balance by conducting sessions 2 and 4 online.

\section{Conclusion and Future Work}
We present novel research regarding the usability of long user-generated passphrases. 
%
We have shown that, when given instruction, users are able to create memorable long passphrases, which potentially could offer more security than their shorter counterparts. Our user study results suggest that our proposed set of policies and recommendations for long passphrases do not compromise usability. 
We also identify two major pitfalls which led to memorability issues: the use of easily interchangeable words and the creation of passphrases that lack a proper grammatical structure.
%
%
Overall, 52\% of participants agreed that they could use passphrases created in our passphrase implementation for logging in every day and 71\%  believed their passphrase was more secure than a password. However, only 34\% agreed that they would prefer to use passphrases over passwords. For this reason, we suggest that long passphrases could be useful for select high-security accounts that require login at most once per day.

Future work could focus on rigorous security analysis by incorporating advanced natural language modeling and informal text corpora scraped from Web, in addition to the COCA dataset. Future work might also extend analysis to languages other than English, or incorporate additional non-dictionary words and slang terms. Our post-study questionnaire revealed that many users believed their passphrases could be guessed if the attacker knew their personal information. Future work into targeted attacks could quantify the impact of publicly available personal information on passphrase guessing. Since we found that users who did not follow proper grammatical structure in their passphrases suffered usability issues, future studies can be conducted on the creation of policies emphasizing proper structure as a requirement. 

\bibliographystyle{IEEEtran}
\bibliography{References.bib}
\end{document}